\documentclass[conference]{IEEEtran}

\usepackage{cite}
\usepackage{amsmath,amssymb}
\usepackage{booktabs}
\usepackage{graphicx}
\usepackage{multirow}
\usepackage{url}
\usepackage[hidelinks]{hyperref}
\usepackage{tikz}
\usetikzlibrary{arrows.meta,positioning,fit}

\title{Frame-Aligned Fusion of Canary and WavLM for Non-Intrusive Intelligibility Prediction of Hearing-Aid-Processed Speech}

\author{
\IEEEauthorblockN{Kazushi Nakazawa}
\IEEEauthorblockA{\textit{Advanced Media, Inc.}\\
Tokyo, Japan}
}

\begin{document}
\bstctlcite{BSTcontrol}
\maketitle

\begin{abstract}
Non-intrusive intelligibility prediction estimates how well hearing-impaired listeners understand hearing-aid-processed speech without a clean reference. We study this task in the 3rd Clarity Prediction Challenge using two frozen speech encoders, Canary and WavLM. The central question is not only whether complementary pretrained representations should be combined, but where their interaction should occur. We compare single-backbone baselines, uniform score averaging, pool-late fusion, cross-attention, frame-aligned fusion, and reverse alignment under a shared left/right-preserving binaural framework. Among the compared systems, the best model temporally prepares WavLM with a learnable strided convolution and fuses it with Canary on the coarser Canary timeline before pooling, reaching Eval RMSE 24.96$\pm$0.06 and Eval Corr 0.796$\pm$0.001. Severity, enhancement-system, layer-window, and temporal-shift analyses indicate that coarse local temporal correspondence before pooling is a useful inductive bias for this task.
\end{abstract}


\section{Introduction}
Predicting the intelligibility of hearing-aid-processed speech is a spoken language technology problem with direct relevance to assistive hearing. A reliable predictor can help evaluate hearing-aid processing systems by estimating the percentage of words that hearing-impaired listeners are expected to understand. The 3rd Clarity Prediction Challenge (CPC3) targets this setting by asking systems to predict sentence-level intelligibility from hearing-aid outputs \cite{barker25_clarity,cpc3_dataset}. In the non-intrusive condition considered here, no clean reference signal is available at inference time.

Recent Clarity-style systems increasingly use pretrained speech models, binaural architectures, metadata, or multi-branch predictors \cite{zezario22_interspeech,cuervo23_clarity,cuervo24_icassp,mogridge24_icassp,zezario24_interspeech,lin25_clarity,buragohain25_clarity,zezario25_clarity}. These results show that speech representations pretrained on large corpora are useful, but they do not fully answer a design question that matters when multiple encoders are available: should different backbones be combined only after utterance-level pooling, or should they interact at the frame level before the sentence-level prediction is formed?

We study this question with two frozen encoders that have different inductive biases. WavLM is a self-supervised model designed for full-stack speech processing and provides relatively fine acoustic-phonetic frame representations \cite{chen2022wavlm}. Canary is an ASR-oriented encoder-decoder model whose internal states are more linguistically organized and operate on a coarser time axis \cite{puvvada2024canary}. If these representations are complementary, the benefit of combining them should depend on the temporal axis and the stage at which interaction is introduced.

This paper makes three contributions. First, it presents a controlled comparison of Canary-only, WavLM-only, uniform score averaging, pool-late fusion, frame-aligned fusion, cross-attention fusion, and reverse alignment under a left/right-preserving binaural framework. Second, it shows that ear-wise frame-aligned fusion, especially with learnable convolutional temporal preparation of WavLM features, outperforms uniform score averaging and utterance-level fusion while using a compact trainable head. Third, it provides severity-wise, enhancement-system-wise, layer-window, and temporal-shift analyses showing that the gain is better explained by coarse local temporal correspondence before pooling than by strict frame synchrony or uniform scalar ensembling alone.

\section{Related Work}
Objective intelligibility metrics such as STOI, ESTOI, MBSTOI, and HASPI are influential baselines for speech and hearing-aid assessment \cite{taal2011stoi,jensen2016estoi,andersen2018mbstoi,kates2021haspi}. They are largely intrusive: a degraded signal is compared with a clean reference. CPC3 instead emphasizes prediction from the processed signal itself, making it a useful benchmark for single-ended models that can be deployed when a clean reference is unavailable.

The Clarity Prediction Challenge series has encouraged learned intelligibility predictors for hearing-aid scenarios \cite{barker22_interspeech,akeroyd24_icassp,barker25_clarity}. Prior systems have explored multi-branch binaural models \cite{zezario22_interspeech}, temporal-hierarchical features from foundation models \cite{cuervo23_clarity}, comparisons of speech foundation models \cite{cuervo24_icassp}, intermediate ASR representations \cite{mogridge24_icassp}, Whisper-based representations and metadata \cite{zezario24_interspeech,radford23a}, and recent CPC3-specific feature-fusion or multi-stage training strategies \cite{lin25_clarity,buragohain25_clarity,zezario25_clarity,yu25_clarity}. Our work is complementary: rather than proposing another single feature source, it isolates how two frozen backbones should interact.

Using frozen encoders with lightweight downstream heads is common in speech representation learning and has been systematized in benchmarks such as SUPERB \cite{yang2021superb}. Self-supervised models such as wav2vec 2.0 and WavLM have shown that large pretrained speech encoders can transfer to diverse tasks \cite{baevski2020wav2vec2,chen2022wavlm}. For intelligibility prediction, this paradigm is attractive because listener-response data are much smaller than pretraining corpora. Binaural interaction also remains important in hearing-aid scenarios, as emphasized by recent non-intrusive binaural predictors \cite{yamamoto25_interspeech}. We therefore preserve left and right channels through most of the trainable stack and compare fusion timing under the same downstream predictor.

Multi-encoder fusion has also been studied in speech assessment and audio-language modeling, where heterogeneous encoders are combined through temporal compression, attention, or adapter modules \cite{wang2026urgentmos,tang2023salmonn,ghosh2024gama,grinberg2026alarm}. These studies motivate the use of complementary representations, but they do not isolate the timing and temporal axis of fusion for hearing-aid intelligibility prediction. Our contribution is a narrower, task-specific study of whether complementary frozen encoders should interact before or after sentence-level pooling.

\section{Method}
\begin{figure*}[t]
\centering
\includegraphics[width=.94\textwidth]{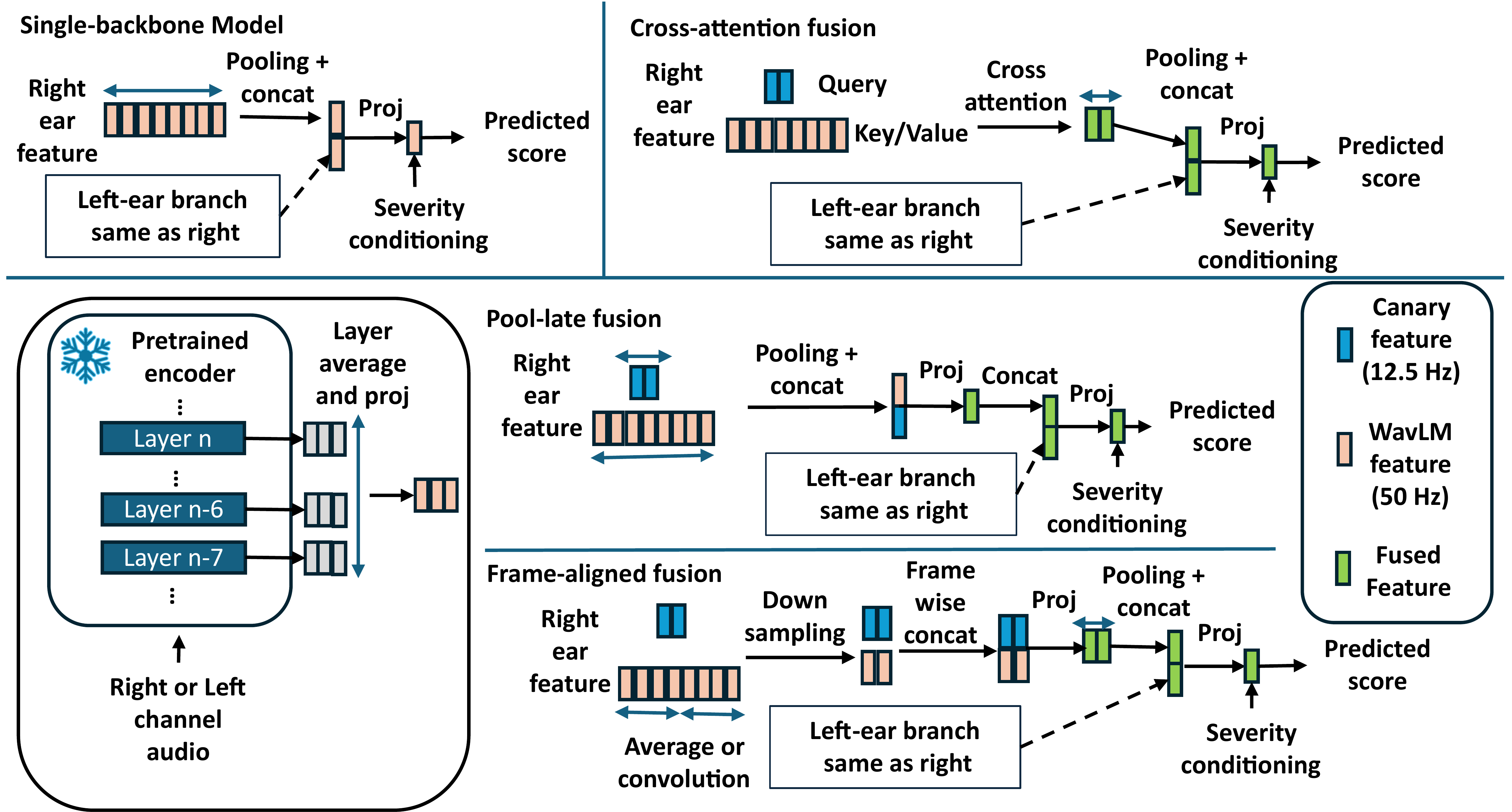}
\caption{Architecture overview. The lower-left inset shows frozen-encoder feature extraction for each right or left channel. Blue, orange, and green bars denote Canary at 12.5~Hz, WavLM at 50~Hz, and fused features, respectively. Pool-late fusion combines pooled utterance vectors, whereas frame-aligned fusion downsamples WavLM and concatenates features frame-wise before pooling. Cross-attention uses Canary queries and WavLM keys/values. Dashed arrows denote the corresponding left-ear branch.}
\label{fig:overview}
\end{figure*}

Figure~\ref{fig:overview} shows the architecture family used in this comparison. The lower-left inset gives the shared feature-extraction path for each right or left channel, while the other panels contrast the single-backbone template, pool-late fusion, frame-aligned fusion, and cross-attention. The figure highlights the two variables tested below: whether Canary--WavLM interaction occurs before or after pooling, and whether binaural information is merged only after the ear-wise representations have been formed.

\subsection{Task and Input Features}
Given a binaural hearing-aid-processed utterance $x$, the objective is to predict an intelligibility score $y\in[0,100]$, corresponding to the percentage of words correctly identified by a hearing-impaired listener \cite{barker25_clarity}. All systems are non-intrusive. Signals are resampled from 32~kHz to 16~kHz, processed as full utterances, padded within minibatches, and accompanied by validity masks so that padding does not affect temporal preparation, attention, pooling, or the loss.

We use frozen Canary (\texttt{nvidia/canary-1b-flash}) and WavLM Large (\texttt{microsoft/wavlm-large}) encoders. Canary layers 10--17 are fixed throughout. The main systems use WavLM layers 17--24 with mean layer aggregation; a separate WavLM-only sweep evaluates layers 5--12, 9--16, 13--20, and 17--24. Both encoders produce 1024-dimensional frame-level representations. For ear $e\in\{L,R\}$, projected features are
\begin{equation}
\mathbf{h}^{(c)}_{e,t_c}=W^{(c)}\mathbf{c}^{(c)}_{e,t_c}, \qquad
\mathbf{h}^{(w)}_{e,t_w}=W^{(w)}\mathbf{c}^{(w)}_{e,t_w},
\end{equation}
where $c$ and $w$ denote Canary and WavLM. The hidden dimension is $d=256$ for single-backbone baselines and $d=192$ for dual-backbone systems.

\subsection{Fusion Strategies}
The comparison preserves the two ears until a late left/right concat-projection. In single-backbone systems, the left and right sequences are pooled separately into $\mathbf{z}_L$ and $\mathbf{z}_R$, then merged as $W_{lr}[\mathbf{z}_L;\mathbf{z}_R]+\mathbf{b}_{lr}$. This delayed merge is intentionally conservative: it avoids introducing a specialized spatial model while preventing the two ears from being averaged before the predictor can use asymmetries between them. The uniform score-averaging control averages predictions from independently trained Canary-only and WavLM-only systems. It is an important baseline because, if the gain from two backbones were mostly scalar variance reduction, this simple control should approach the best learned fusion model. It should be interpreted as a deliberately simple scalar ensembling control rather than an exhaustive score-fusion baseline.

Trainable parameter counts are reported without frozen backbones. The single-backbone baselines use $d=256$, while dual-backbone systems use $d=192$ and the same one-layer residual trunk. This is not a fully compute-matched study because frozen encoders differ, but it keeps the trainable prediction heads compact and makes it possible to test whether the proposed inductive bias improves performance without simply increasing downstream capacity.

The compared systems differ only in the stage at which backbone and ear information are combined. Canary-only and WavLM-only test each representation under the same single-backbone template. Uniform score averaging tests scalar-level complementarity under a fixed one-half mixture weight. Pool-late fusion tests utterance-level representation complementarity. Frame-aligned fusion tests local pre-pooling complementarity after explicit temporal preparation. Cross-attention tests flexible sequence-level interaction without an explicit local alignment bias. Reverse alignment tests whether the same pre-pooling idea remains effective when the coarser Canary sequence is expanded toward the WavLM time axis.

Pool-late fusion first pools Canary and WavLM within each ear, then fuses the utterance-level vectors:
\begin{equation}
\mathbf{z}^{(f)}_e = W_{late}[\mathbf{z}^{(c)}_e;\mathbf{z}^{(w)}_e]+\mathbf{b}_{late}.
\end{equation}
This tests whether cross-backbone complementarity remains useful after both representations have already been reduced to utterance-level summaries.

Frame-aligned fusion instead introduces interaction before pooling. Let $H^{(c)}_e\in\mathbb{R}^{T_c\times d}$ and $H^{(w)}_e\in\mathbb{R}^{T_w\times d}$ be the projected sequences. Since WavLM operates at a higher frame rate, we temporally prepare it with a fixed masked average-downsampling module or a learnable one-dimensional convolution with kernel size 4 and stride 4. The prepared WavLM sequence is then adaptively mapped to the Canary length and fused on the Canary timeline:
\begin{equation}
\bar{H}^{(w)}_e=\mathcal{A}(\mathcal{D}(H^{(w)}_e),T_c),\quad
H^{(f)}_{e,t}=W_f[H^{(c)}_{e,t};\bar{H}^{(w)}_{e,t}]+\mathbf{b}_f.
\end{equation}
The resulting left and right fused sequences are merged at the sequence level before pooling. The downsampling factor is four because the selected WavLM representation is approximately four times denser than the selected Canary representation. The fixed average path tests whether simple local summarization is sufficient. The convolutional path tests whether the model benefits from learning how to summarize fine-rate WavLM evidence before it is compared with Canary states.

Reverse frame-aligned variants map Canary upward to the WavLM timeline using linear interpolation or transposed convolution. These variants test whether the benefit comes from frame-level interaction in general or from choosing Canary as the reference time axis. We also test cross-attention in the standard query-key-value form \cite{vaswani2017attention}, with Canary as query and WavLM as key-value; a reverse variant uses WavLM as query. Cross-attention provides a flexible non-local interaction mechanism, but it does not impose the same local temporal correspondence as frame-aligned fusion.

\subsection{Pooling, Conditioning, and Training}
The downstream stack is fixed within each comparison family. A residual temporal convolution is followed by a one-layer bidirectional LSTM and additive attention pooling \cite{hochreiter1997lstm,bahdanau2015neural}. With encoded sequence $\bar{\mathbf{f}}_{1:T}$ and mask $m_t$, attention pooling is
\begin{equation}
\begin{aligned}
e_t &= \mathbf{w}^{\top}\tanh(W_a\bar{\mathbf{f}}_t),\\
\alpha_t &= \frac{m_t\exp(e_t)}{\sum_u m_u\exp(e_u)},\\
\mathbf{u} &= \sum_t \alpha_t\bar{\mathbf{f}}_t .
\end{aligned}
\end{equation}
A residual MLP trunk follows. A listener-severity label is inserted late through a learned embedding and low-rank adapter. The scalar prediction is bounded by $\hat{y}=100\sigma(r)$, and training minimizes mean squared error on normalized targets.

All trainable modules are optimized with AdamW \cite{kingma2015adam,loshchilov2019decoupled} using learning rate $10^{-4}$, weight decay $10^{-3}$, batch size 64, and gradient clipping at 1.0 \cite{pascanu2013difficulty}. Backbone features are cached as selected-layer means so that all variants operate on identical frozen inputs. Models are trained for five epochs and the checkpoint with the lowest validation RMSE is selected.

\subsection{Temporal-Shift Control}
To test whether frame-aligned fusion depends on genuine temporal correspondence rather than only on additional trainable capacity, we shift the temporally prepared WavLM sequence before fusion in the best model. The shift values are $\Delta\in\{-4,-2,-1,0,1,2,4\}$ steps, where one step corresponds to 80~ms after preparation. Vacated regions are zero-padded and marked invalid rather than wrapped. This control asks whether performance collapses under modest misalignment and whether exact zero-offset synchrony is necessary. To save space, the results table reports representative offsets from this sweep.

\section{Experimental Setup}
All experiments use CPC3 \cite{barker25_clarity,cpc3_dataset}. The official development and evaluation sets are fixed. On the official training split, we perform 5-fold cross-validation grouped by scene token to reduce leakage across related items. For each seed in $\{1,2,3,4,5\}$, one model is trained per fold and predictions are averaged on the official development and evaluation sets. Model weights and checkpoints are fit on training-fold validation data; the evaluation set is used only for held-out reporting of the fixed analysis variants in this manuscript. Results are reported as mean and standard deviation over the five seed-level ensembles. These deviations are not item-level confidence intervals and are not a substitute for paired significance testing. The uniform score-averaging system is produced from the matched Canary-only and WavLM-only systems rather than from a separately tuned score-level regressor. Following CPC3, we report root-mean-square error (RMSE) and Pearson correlation (Corr); lower RMSE and higher Corr are better. For group analyses, MAE is also reported because it makes calibration changes easier to interpret than squared-error metrics alone.

\begin{table*}[t]
\centering
\caption{Main comparison under the left/right-preserving setting. Values are mean$\pm$standard deviation across five seed-level ensembles. Parameter counts exclude frozen backbones; the uniform score-averaging row reports the sum of the two constituent single-backbone predictors.}
\label{tab:main_results}
\scriptsize
\setlength{\tabcolsep}{3.8pt}
\begin{tabular}{lccccccc}
\toprule
System & WavLM layers & Prep. & Params & Dev RMSE & Dev Corr & Eval RMSE & Eval Corr \\
\midrule
Canary-only baseline & -- & -- & 1.60M & 22.75$\pm$0.32 & 0.827$\pm$0.004 & 25.64$\pm$0.14 & 0.784$\pm$0.002 \\
WavLM-only baseline & 17--24 & -- & 1.60M & 24.57$\pm$0.28 & 0.800$\pm$0.002 & 26.62$\pm$0.13 & 0.766$\pm$0.002 \\
Uniform score avg. & 17--24 & -- & 3.20M & 23.26$\pm$0.19 & 0.818$\pm$0.003 & 25.53$\pm$0.15 & 0.784$\pm$0.003 \\
Pool-late fusion & 17--24 & -- & 1.69M & 22.77$\pm$0.33 & 0.828$\pm$0.004 & 25.57$\pm$0.10 & 0.786$\pm$0.002 \\
Frame-aligned fusion & 17--24 & Avg & 1.15M & 22.65$\pm$0.17 & 0.827$\pm$0.003 & 25.03$\pm$0.06 & 0.794$\pm$0.001 \\
Frame-aligned fusion & 17--24 & Conv & 1.30M & \textbf{22.52$\pm$0.14} & \textbf{0.829$\pm$0.002} & \textbf{24.96$\pm$0.06} & \textbf{0.796$\pm$0.001} \\
Cross-attention fusion & 17--24 & -- & 1.52M & 22.89$\pm$0.32 & 0.824$\pm$0.004 & 25.62$\pm$0.21 & 0.785$\pm$0.003 \\
\bottomrule
\end{tabular}
\end{table*}

\begin{figure}[t]
\centering
\includegraphics[width=\linewidth]{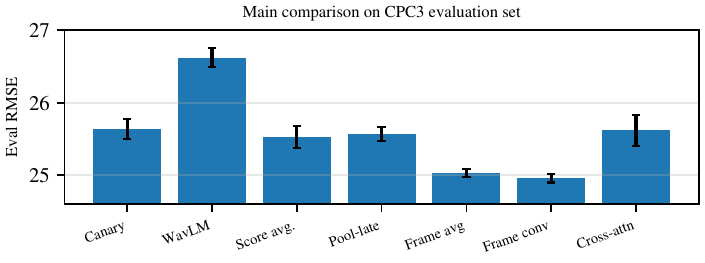}
\caption{Eval RMSE for the main systems. Error bars show standard deviation across five seed-level ensembles. Frame-aligned fusion with convolutional temporal preparation is the strongest system among the compared variants; small within-family differences should be interpreted with the seed-level variability in mind.}
\label{fig:eval_rmse}
\end{figure}

\section{Results and Analysis}
\subsection{Main Comparison}
Table~\ref{tab:main_results} shows three main patterns. First, the Canary-only baseline is strong, while the WavLM-only baseline is substantially weaker. Second, uniform score averaging improves Eval RMSE over Canary-only by only 0.11 and does not improve Eval Corr, indicating that output-level ensembling alone does not fully exploit the second backbone. Third, frame-aligned fusion is the strongest dual-backbone family among the compared systems. With convolutional temporal preparation, it improves Eval RMSE by 0.69 over Canary-only and by 0.58 over uniform score averaging, while also improving Eval Corr by about 0.012. Because no item-level paired significance test is included, the small Avg--Conv gap should be treated as a within-family preference rather than as a decisive separation.

The best system is not the largest trainable model: it uses 1.30M trainable parameters, fewer than the Canary-only baseline, pool-late fusion, and cross-attention fusion, and far fewer than the two-model uniform score average. Because frozen-encoder compute and hidden dimensions are not perfectly matched, this is not a definitive capacity study. Nevertheless, the comparison makes a simple explanation unlikely: the gain is associated with temporally prepared pre-pooling interaction, not merely with adding a second feature stream or a larger predictor.

For scale, the reported CPC3 Eval results place strong systems in a similar absolute error range \cite{clarity2025_results}. Because our model was evaluated offline rather than submitted through the official CPC3 procedure, this statement is intended only as context and should not be interpreted as a leaderboard claim or an official rank.

\begin{table*}[t]
\centering
\caption{Diagnostic analyses. Panel A tests reverse alignment. Panel B sweeps WavLM-only layer windows. Panel C reports selected shifts of the temporally prepared WavLM stream in the best frame-aligned system.}
\label{tab:diagnostic}
\scriptsize
\setlength{\tabcolsep}{3.5pt}
\begin{minipage}{0.33\textwidth}
\centering
\textbf{A. Reverse alignment}\\[0.8mm]
\begin{tabular}{lccc}
\toprule
Method & Eval RMSE & Corr & MAE \\
\midrule
Canary-up, linear & \textbf{25.26} & \textbf{0.791} & \textbf{17.84} \\
Canary-up, transp. conv. & 25.46 & 0.788 & 18.05 \\
Reverse cross-attn. & 25.63 & 0.785 & 18.10 \\
\bottomrule
\end{tabular}
\end{minipage}
\hfill
\begin{minipage}{0.31\textwidth}
\centering
\textbf{B. WavLM-only layers}\\[0.8mm]
\begin{tabular}{lcc}
\toprule
Layers & Eval RMSE & Corr \\
\midrule
5--12 & 28.12 & 0.743 \\
9--16 & 27.26 & 0.756 \\
13--20 & 26.70 & \textbf{0.767} \\
17--24 & \textbf{26.62} & 0.766 \\
\bottomrule
\end{tabular}
\end{minipage}
\hfill
\begin{minipage}{0.31\textwidth}
\centering
\textbf{C. Temporal shift}\\[0.8mm]
\begin{tabular}{rcc}
\toprule
Shift & Eval RMSE & Corr \\
\midrule
-320 ms & 25.12 & 0.793 \\
-80 ms & 24.99 & 0.795 \\
0 ms & 24.96 & 0.796 \\
+160 ms & \textbf{24.93} & \textbf{0.796} \\
+320 ms & 24.96 & 0.796 \\
\bottomrule
\end{tabular}
\end{minipage}
\end{table*}

\subsection{Temporal Axis, Layer Choice, and Shift Controls}
Panel A of Table~\ref{tab:diagnostic} shows that reverse alignment is useful but weaker than the proposed WavLM-to-Canary direction. Mapping Canary upward to the WavLM timeline with linear interpolation reaches Eval RMSE 25.26, which is better than several non-frame-aligned controls in Table~\ref{tab:main_results}. However, it remains 0.30 RMSE worse than WavLM-to-Canary convolutional preparation. Thus, frame-level interaction helps in both directions, but the coarser Canary timeline is the better reference axis under the present architecture and optimization recipe.

Panel B shows that upper WavLM layers are the most useful for the WavLM-only baseline: layers 17--24 give the lowest Eval RMSE, and layers 13--20 give the highest Eval Corr. This supports the choice of layers 17--24 for the main dual-backbone comparison. Panel C refines the interpretation of ``alignment.'' Exact zero shift is competitive, but the central region from approximately -80 to +160~ms is very flat, and +160~ms slightly improves mean Eval RMSE. The result should therefore not be read as evidence for strict frame synchrony. A more defensible interpretation is that the model benefits from preserving coarse local temporal correspondence while remaining tolerant to modest residual offsets introduced by feature extraction, subsampling, and downstream pooling.

\begin{table*}[t]
\centering
\caption{Robustness analyses on the official evaluation set. Severity rows compare Canary-only with the best frame-aligned system. The system-wise panel reports macro averages over nine enhancement systems and win rates against Canary-only.}
\label{tab:robustness}
\scriptsize
\setlength{\tabcolsep}{3.5pt}
\begin{minipage}{0.58\textwidth}
\centering
\textbf{A. Listener severity}\\[0.8mm]
\begin{tabular}{l l c c c c}
\toprule
System & Severity & $N$ & RMSE & Corr & MAE \\
\midrule
\multirow{3}{*}{Canary-only}
& Mild & 2340 & 24.75 & 0.772 & 17.32 \\
& Moderate & 4908 & 25.98 & 0.783 & 18.43 \\
& Mod.-severe & 426 & 26.52 & 0.789 & 19.34 \\
\midrule
\multirow{3}{*}{Frame-aligned, Conv}
& Mild & 2340 & \textbf{24.20} & \textbf{0.783} & \textbf{17.09} \\
& Moderate & 4908 & \textbf{25.26} & \textbf{0.793} & \textbf{18.29} \\
& Mod.-severe & 426 & \textbf{25.48} & \textbf{0.806} & \textbf{18.69} \\
\bottomrule
\end{tabular}
\end{minipage}
\hfill
\begin{minipage}{0.39\textwidth}
\centering
\textbf{B. Enhancement-system macro summary}\\[0.8mm]
\begin{tabular}{lccc}
\toprule
System & RMSE & Corr & MAE \\
\midrule
Canary-only & 24.57 & 0.641 & 17.99 \\
Uniform score avg. & 24.43 & 0.647 & 18.85 \\
Frame-align. & \textbf{23.95} & \textbf{0.661} & \textbf{17.80} \\
\bottomrule
\end{tabular}

\vspace{1mm}
\begin{tabular}{lccc}
\toprule
System & RMSE win & Corr win & MAE win \\
\midrule
Uniform score avg. & 5/9 & 6/9 & 0/9 \\
Frame-align. & \textbf{9/9} & \textbf{9/9} & \textbf{6/9} \\
\bottomrule
\end{tabular}
\end{minipage}
\end{table*}

\subsection{Severity and Enhancement-System Analyses}
Table~\ref{tab:robustness} shows that the aggregate gain is not concentrated in a single listener group or enhancement system. Frame-aligned fusion improves RMSE, Corr, and MAE over Canary-only for mild, moderate, and moderately severe listeners. The largest RMSE reduction is observed for the moderately severe group, suggesting that local WavLM-derived acoustic evidence may be especially helpful when intelligibility is more strongly affected by hearing loss. This trend should be interpreted cautiously because that group is smaller than the mild and moderate groups.

The enhancement-system macro summary gives a second view. Uniform score averaging improves RMSE for five of nine systems and Corr for six of nine systems, but it increases macro MAE and wins on MAE for none of the systems. Frame-aligned fusion improves RMSE and Corr for all nine systems and reduces MAE for six. This pattern supports the central claim: Canary and WavLM are complementary, but their complementarity is more reliably exploited through frame-level interaction before sequence pooling than through uniform scalar averaging after two independent predictors have already been formed.

\subsection{What the Ablations Rule Out}
The ablations in Tables~\ref{tab:main_results}--\ref{tab:robustness} rule out several simpler explanations for the gain. First, the improvement is not explained by uniform score averaging: averaging slightly reduces Eval RMSE relative to Canary-only, but it does not improve Eval Corr and is weaker than frame-aligned fusion. Second, the improvement is not merely a consequence of a larger trainable head. The best frame-aligned model has fewer trainable parameters than the single-backbone baselines and less than half the trainable parameters of the two-model uniform score average. Third, unconstrained cross-attention under this compact head is not sufficient. This does not imply that cross-attention is generally unsuitable; rather, without an explicit locality or relative-position bias, it does not provide the same inductive bias as temporal preparation followed by local frame-wise fusion.

A useful way to view the uniform score-averaging control is through calibration. If the two single-backbone predictors have errors $e_c$ and $e_w$, the average reduces squared error only when the WavLM error is not too large and is sufficiently complementary to the Canary error. It also fixes the mixture weight to one half for every listener, sentence, and enhancement system. This is a restrictive assumption for CPC3, where some scenes may be dominated by linguistic recoverability and others by local enhancement artifacts. A learned pre-pooling fusion head can instead condition the use of WavLM evidence on the local Canary state and on the downstream temporal context, which helps explain why the frame-aligned model improves MAE more reliably than uniform score averaging in the system-wise analysis.

The diagnostic experiments further separate frame-level interaction from the choice of reference axis. Reverse alignment improves over several non-frame-aligned controls, so the benefit is not unique to one implementation of downsampling. However, it remains weaker than mapping WavLM to the Canary timeline, suggesting that the coarser ASR-oriented representation is a better anchor for this sentence-level prediction task. The temporal-shift control shows a broad optimum around zero shift rather than a sharp peak. This suggests that the model uses local neighborhoods of acoustic evidence, not exact frame synchrony.

Choosing the Canary axis is also computationally and statistically meaningful. Expanding Canary to the WavLM frame rate repeats or interpolates a coarse representation over many fine frames, increasing sequence length without adding new linguistic evidence. Compressing WavLM first has the opposite effect: it forces local acoustic evidence to be summarized into short neighborhoods before interaction, reduces the number of tokens seen by the recurrent and attention-pooling layers, and provides an explicit regularizer for the amount of cross-stream correspondence the model must learn from limited listener-response labels. The convolutional preparation path keeps this regularization while allowing the summary within each local WavLM neighborhood to be learned rather than fixed.

The WavLM layer sweep reinforces the same interpretation. Lower and middle WavLM layers are less effective for the standalone WavLM-only predictor, while upper layers give the strongest RMSE and correlation. These layers are likely closer to phonetic and word-related evidence than to raw acoustics, which makes them a better complement to Canary. However, the WavLM-only result remains weaker than Canary-only, so the gain in the proposed system should not be interpreted as replacing ASR-oriented features with self-supervised features. The improvement comes from using upper-layer WavLM as an auxiliary local evidence stream attached to a stronger linguistic scaffold.

\section{Discussion and Limitations}
The results favor a constrained-fusion interpretation. Sentence intelligibility is an utterance-level target, but the evidence that changes a listener response can be local: a few masked words or enhancement artifacts may have a large effect. Pool-late fusion discards much of this locality before Canary and WavLM interact. Frame-aligned fusion keeps local structure long enough to compare Canary's coarser linguistic states with WavLM's finer acoustic-phonetic evidence. The convolutional preparation path is useful because it learns how to summarize short WavLM neighborhoods before this comparison, instead of assuming that all fine-rate frames contribute equally.

The group-wise results support this interpretation beyond the aggregate RMSE. Frame-aligned fusion improves all three listener-severity groups and all nine enhancement systems in RMSE and Corr. Its MAE behavior is also more stable than uniform score averaging: the scalar average wins on RMSE for some enhancement systems but never wins on MAE, whereas frame-aligned fusion reduces MAE for six systems. This suggests that pre-pooling fusion gives the regression head a better opportunity to use the second stream selectively, rather than relying on a fixed mixture weight.

Several technical limitations remain. The comparison is controlled in evaluation protocol, frozen features, folds, pooling module, and severity conditioning, but it is not perfectly matched in frozen-encoder compute or hidden dimension. The study uses only two encoders and one challenge dataset, so the preference for the Canary time axis should be treated as specific to this representation pair until tested more broadly. Listener conditioning is limited to severity groups, and the sentence-level target does not identify which words or phonemes drive each prediction error. The score-level baseline is a uniform average rather than a validation-tuned stacking model. Finally, five seed-level ensembles do not replace item-level paired significance testing; bootstrap tests and stronger disruption controls such as temporal permutation or cross-utterance feature replacement would further isolate the role of local correspondence.

\subsection{Future Work}
Future work should test whether the same alignment principle holds with other ASR encoders, enhancement-oriented encoders, or audio-language representations. It should also add validation-tuned score fusion, paired bootstrap testing over evaluation items, dynamic alignment or lightweight adapters, richer binaural interaction before the final left/right merge, and prediction heads conditioned on detailed audiometric information. These studies would clarify whether the present result is specific to Canary and WavLM or reflects a broader principle: use a coarse linguistic stream as the reference axis and inject fine-rate acoustic evidence before pooling.

\section{Conclusion}
We presented a controlled study of non-intrusive intelligibility prediction for hearing-aid-processed speech using frozen Canary and WavLM encoders. Under a left/right-preserving binaural framework, frame-aligned fusion with learnable convolutional temporal preparation performed best among the compared systems, improving Eval RMSE from 25.64 for Canary-only and 25.53 for uniform score averaging to 24.96. Additional analyses show consistent gains across listener-severity groups and enhancement systems, stronger upper-layer WavLM behavior, and an alignment benefit better described as coarse local correspondence than strict synchrony. These findings support temporal preparation and pre-pooling ear-wise interaction as useful design principles for combining complementary frozen speech encoders.

\section*{Acknowledgment}
This manuscript used generative AI for English editing and wording suggestions in the manuscript text. All scientific claims, experiments, and final text were reviewed and validated by the authors, who take responsibility for the submitted manuscript.
\clearpage
\bibliographystyle{IEEEtran}
\bibliography{ref}

\end{document}